\documentclass[conference]{IEEEtran}

\usepackage{graphicx}
\usepackage{latexsym}
\usepackage{amsfonts}
\usepackage{amssymb}
\usepackage{amsbsy}
\usepackage{color}
\usepackage{amsmath}
\usepackage{multicol}
\usepackage{float}
\usepackage{subfigure}
\usepackage{algorithm}
\usepackage{algorithmic}
\usepackage[dvips]{epsfig}

\usepackage{caption}

\usepackage{enumitem}
\usepackage{graphicx}
\usepackage{latexsym}
\usepackage{amsfonts}
\usepackage{amssymb}
\usepackage{amsbsy}
\usepackage{color}
\usepackage{amsmath}
\usepackage{multicol}
\usepackage{float}
\usepackage{subfigure}
\usepackage{algorithm}
\usepackage{algorithmic}
\usepackage[dvips]{epsfig}

\begin{document}

\title{\LARGE Ultra-Dense Edge Caching under \\Spatio-Temporal Demand and Network Dynamics}

\author{\IEEEauthorblockN{Hyesung Kim, Jihong Park$^*$, Mehdi Bennis$^\dagger$, Seong-Lyun Kim, and M\'erouane Debbah$^{\circ}$}
\IEEEauthorblockA{\small
School of Electrical and Electronic Engineering, Yonsei University,
Seoul, Korea, email: \{hskim, slkim\}@ramo.yonsei.ac.kr\\
$^*$Department of Electronic Systems, Aalborg University, Denmark, email: jihong@es.aau.dk\\
$^\dagger$Centre for Wireless Communications, University of Oulu, Finland, email: bennis@ee.oulu.fi\\$^{\circ}$
Mathematical and Algorithmic Sciences Lab, Huawei France R\&D, Paris, France \\
and Large Networks and Systems Group (LANEAS), CentraleSup\'elec, Gif-sur-Yvette, France, email: merouane.debbah@huawei.com
}
}


\maketitle

\begin{abstract}

This paper investigates a cellular edge caching design under an extremely large number of small base stations (SBSs) and users. In this ultra-dense edge caching network (UDCN), SBS-user distances shrink, and each user can request a cached content from multiple SBSs. Unfortunately, the complexity of existing caching controls' mechanisms increases with the number of SBSs, making them inapplicable for solving the fundamental caching problem: How to maximize local caching gain while minimizing the replicated content caching? Furthermore, spatial dynamics of interference is no longer negligible in UDCNs due to the surge in interference. In addition, the caching control should consider temporal dynamics of user demands. 
To overcome such difficulties, we propose a novel caching algorithm weaving together notions of mean-field game theory and stochastic geometry. These enable our caching algorithm to become independent of the number of SBSs and users, while incorporating spatial interference dynamics as well as temporal dynamics of content popularity and storage constraints. Numerical evaluation validates the fact that the proposed algorithm reduces not only the long run average cost by at least 24\% but also the number of replicated content by 56\% compared to a popularity-based algorithm. 


\end{abstract}


\section{Introduction}


Upcoming 5G systems are expected to become ultra-dense networks (UDNs) due to relentless user demand growth \cite{UDN_sur}, \cite{MF_ref1}. Limited backhaul capacity is however unable to cope with a huge number of small base stations (SBSs) in a UDN. In this respect, edge caching is a promising UDN enabler that alleviates backhaul congestion by storing popular content during off-peak hours at the network edges, i.e. SBSs \cite{Living}-- \cite{caching tradeoff: 2016}, yielding to an \emph{ultra-dense edge caching network (UDCN)}. This paper aims at proposing a UDCN caching control algorithm by solving the following three fundamental issues in UDCNs.

\textbf{1. Caching control complexity}. 
Edge caching network performance relies on how much amount of users request the content cached in SBSs, i.e. cache hits. If user demand is fully correlated, caching the most popular files would be an optimal control policy. However, user demand in reality is heterogeneous, and therefore leads to the fundamental trade-off in the edge caching: maximizing individual user's cache hits while minimizing replicated content caching at SBSs. Solving this trade-off in a centralized manner is too complicated to be implemented in practice, so preceding works suggest semi-distributed caching control algorithms where each SBS only exploits the local caching decisions of neighboring SBSs \cite{caching_local_info}. Unfortunately, the number of neighboring SBSs in a UDCN is much larger than that of a traditional network, and thus existing algorithms are computationally infeasible under ultra-dense SBS environment. To make these worse, the following two spatio-temporal aspects of UDCNs further increase the caching control complexity.

  \begin{figure}
\centering
\includegraphics[angle=0, height=0.4\textwidth]{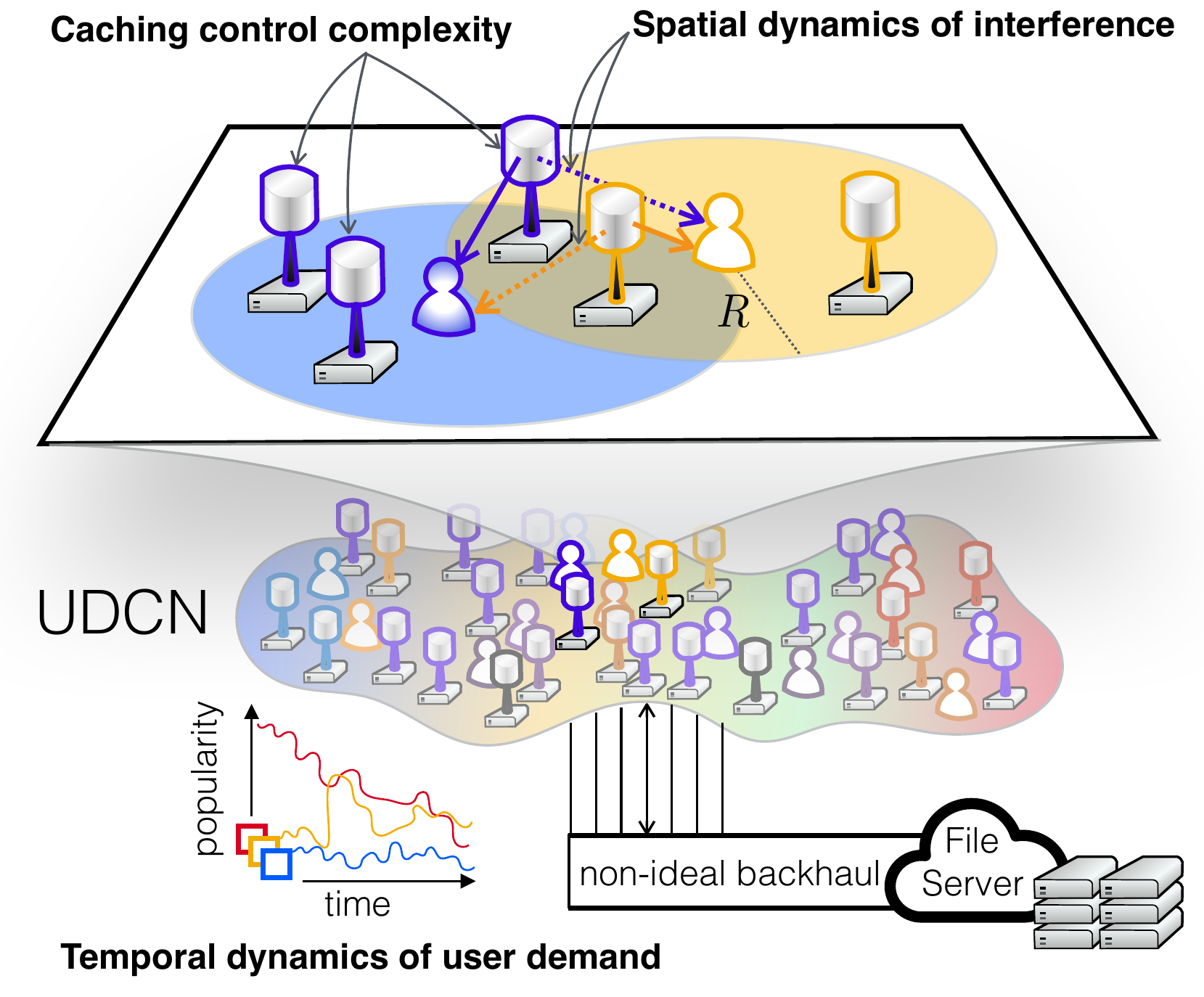}   
\caption{\small{An illustration of ultra-dense edge caching networks (UDCNs) with spatio-temproal user demand and network dynamics.  }}\label{system_model} \vskip -10pt
\end{figure}

\textbf{2. Temporal dynamics of user demand}. 
Content popularity among users evolves over time, so demand prediction error exists when caching files at SBSs. Minimizing this prediction error straightforwardly brings about huge complexity increase, but nevertheless UDCN caching control cannot overlook this misprediction impact. The reason is that numerous users amplify individual misprediction loss in a UDCN. A large number of SBSs also lead to significant redundant caching due to misprediction, increasing congestion at both backhauls and cache storages.

\textbf{3. Spatial dynamics of interference}. In the downlink, interference amount and variance increase linearly with the number of SBSs \cite{interf_var}. This significant impact of interference should thus be carefully considered in UDCN caching control design. Many preceding works apply somewhat simplified interference modelings such as the average interference \cite{caching_avg_interf} with fixed SBS locations. Incorporating the interference distribution that specifies random locations of SBSs however incurs additional complexity in UDCN caching control.

On the basis of these issues, our objective is to propose a UDCN caching control algorithm that incorporates spatio-temporal demand, cache storage and interference dynamics with low computational complexity while minimizing the long run average (LRA) cost. The cost is defined such that it  increases with (i) backhaul use, (ii) cache storage use, and (iii) content replication, while decreasing with (iv) spectral efficiency.

To this end, we tackle the aforementioned triple difficulty in reverse order by using stochastic geometry (SG) and mean-field game (MFG) theory. SG first plays a key role to specify the interference dynamics at a randomly selected user, i.e. a typical user. In UDNs, the interference normalized by BS densification becomes deterministic \cite{JHP1}, \cite{JHP2}, which allows us to simplify the spatial dynamics of UDCNs. Based on this result, MFG captures temporal dynamics of demand and its corresponding caching behavior by using two partial differential equations (PDEs), Hamilton-Jacobi-Bellman equation (HJB) and Fokker-Planck-Kolmogorov equation (FPK). In a UDCN, remarkably, no matter how many SBSs and users exist, only a single pair of HJB-FPK can describe the entire caching control and the resultant LRA cost dynamics \cite{MFG_application}, achieving fixed computational complexity. Solving these HJB and FPK equations provides the desired optimal UDCN caching control.

%
%

\textbf{Related works}.  A recent study on caching \cite{caching_dyna2} considers interference dynamics using SG for the spatial domain. Caching policy adaptive to time-varying user request \cite{caching_dyna1} is proposed to enhance the local caching gain.
 The authors of \cite{MFG_caching} considered temporal dynamics of SBSs' states, replicated contents, and interference on caching control in UDN environment. 
They, however, neglected temporal evolution of content popularity and spatial dynamics of interference determined by user locations. 
 None of the works on caching jointly considers the spatio-temproal dynamics of user demand and network (interference and SBSs' backhaul and storage capacities) in ultra-dense scenarios. 
 
Under these spatio-temporal dynamics, it is hard to improve the caching gain while minimizing the replicated content caching  \cite{social_cost1}. 
The replication problem becomes more severe in UDCN, since the increased number of candidate SBSs serving the same user induces that only one of them transmits the cached data. Furthermore, the amount of replicated contents varies according to the aforementioned dynamics. A caching strategy dealing with the dynamically evolving replication remains an open problem in UDNs.

\textbf{Contributions}. The contributions of this paper are summarized as below.
\begin{itemize} 
\item A novel caching control algorithm for UDCNs is proposed, which minimizes LRA cost while incorporating spatio-temporal demand and interference dynamics with fixed computational complexity, independent of the numbers of SBSs and users (see Proposition 1). Specifically, the algorithm does not need to get full knowledge of other SBSs' caching strategies or storage states.
\item 
The proposed algorithm's gains in LRA cost and replicated contents are numerically validated (Figs. 5 and 6).

\item The unique mean-field equilibrium (MFE) \cite{MFG_application} is verified by numerical evaluation (Fig. 4) as well as by analysis (Proposition 1).


\end{itemize}

\section{Network model and spatio-temporal dynamics}

\subsection{Ultra-dense network}

 We consider a UDN with SBS density $\lambda_b$ and user density $\lambda_u$. Their locations follow independent homogeneous Poisson point processes (PPPs), respectively. 
We set $\lambda_b \gg \lambda_u$, according to the definition of UDN \cite{interf_udn}.
User $i$ receives signals from SBSs within a reception ball $b(y_i,R)$ centered at $y_i$ with radius $R$, which represents the average distance
that determines a region where the received signal power is larger than noise floor, as depicted in Fig. 1. When $R$ goes to infinity, the reception ball model is identical to a conventional PPP network.  

Transmitted signals from SBSs experience path-loss attenuation. The attenuation from the $k$-th SBS coordinates $z_k$ to the $i$-th user coordinates $y_i$
is  $l_{k,i}=\min(1,||z_k-y_i||^{-\alpha})$, where $\alpha$ is the path-loss exponent.
The transmitted signals experience independent and identically distributed fading with the coefficient $g_{k,i}(t)$. We assume that the coefficient is not temporally correlated. The channel gain $h_{k,i}$ from SBS $k$ to user $i$ is  $|h_{k,i}(t)|^2=l_{ki}|g_{k,i}(t)|^2$. The received signal power is given as $S(t)=P|h_{k,i}(t)|^2$, where $P$ denotes transmit power of an SBS.

The SBS directionally transmits signal by using $N_a$ number of antennas. Its beam pattern at a receiver follows a sectored uniform linear array model \cite{MIMO} where the main lobe gain is $N_a$ with beam width $\theta_{N_a} = 2\pi/\sqrt{N_a}$ assuming that side lobes are neglected and the beam center points at the receiver.

\subsection{Edge caching model}
Let us assume that there is a set $\mathcal{N}$ consisting of $N$ SBSs within the reception ball with radius $R$.
Each SBS $k\in\mathcal{N}$ has data storage unit size $C_{k,j}$ assigned for  content $j$.
The storage units allow SBSs to download contents a priori from a server connected with non-ideal backhaul link as depicted in Fig. \ref{system_model}. 
We assume that there is a set $\mathcal{M}$ consisting of $M$ contents, and the server has all the contents files in the set $\mathcal{M}$.
Users request  content $j$
from the contents set $\mathcal{M}$ with probability $x_j$, and the size of the content $j$ is denoted by $L_j$. 
The goal of SBS $k$ is to determine a vector of caching probabilities $\boldsymbol{p}_k(t)=\{p_{k,1}(t),...,p_{k,j}(t),...p_{k,M}(t)\}$, 
where $p_{k,j}(t) \in [0,1]$ is a probability that SBS $k$ downloads content $j$ through its own backhaul link at time $t$.
The control variable $p_{k,j}(t)$ can also be interpreted as a fraction of the file in case that each content file is encoded using a maximum distance separable dateless code \cite{MDS}. 

According to the definition of UDN, multiple SBSs can serve a user \cite{interf_udn} unlike traditional unidentified networks as show in Fig. 1.
We assume that a reference user is associated to one of the SBSs storing the requested content within the threshold distance $R$. If there are multiple SBSs having the same content within the region, the serving SBS is randomly chosen.
If no SBS has cached the requested content, one of the SBSs within $R$ fulfills the user request by downloading it from the server through the backhaul.




\subsection{ Spatio-temporal dynamics of user demand and network}

\textbf{Evolution law of time-varying content popularity.} The probability that users request a content, representing popularity, varies over time in reality.  We use Ornstein-Uhlenbeck process \cite{MF_ref2} to model this temporal fluctuation as a stochastic differential equation (SDE):
\vskip -10pt
\begin{equation}
\text{d}x_j(t)= (u_j-a_j)\text{d}t+\eta \text{d}W_j(t), \label{SDE_x}
\end{equation}    \vskip-5pt
\noindent where $x_j(t)$ is the probability that users request content $j$ at time $t$, $u_j$ and $a_j$ respectively denote popularity increment and decrement for content $j$, $\eta$ is a positive constant, and $W_j(t)$ is a Wiener process. 




\textbf{Temporal dynamics of cache storage size.} The remaining cache storage capacity varies according to the instantaneous caching policy. 
Let us assume that SBS $k$ discards content files at a rate of $\mu_{k,j}$
in order to make a space for downloading other data.
Considering the discarding rate, we model
the evolution law of storage unit as follows \vskip -10pt
\begin{equation}
\text{d}Q_{k,j}(t)=(\mu_{k,j}-L_j p_{k,j}(t))\text{d}t, \label{SDE_q}
\end{equation}
where $Q_{k,j}(t)$ denotes the remaining storage size dedicated to content $j$ of SBS $k$ at time $t$, and $L_j$ is data size of content~$j$. $L_jp_{k,j}(t)$ represents instant data size of content $j$ downloaded by SBS $k$ at time $t$.

%
%
%

\textbf{Spatial dynamics of interference}. In UDN, an SBS having no associated user within its coverage becomes dormant, not transmitting any signal. Locations of users determine activation of SBSs, characterizing spatial distribution of interference.  
Consider a randomly selected typical user. 
We assume that active SBSs have always data to transmit regardless of their own caching policy. It indicates that 
the aggregate interference is imposed by active SBSs with active probability $p_a$. Assuming that $p_a$ is homogeneous over SBSs yields $p_a \approx 1-[1+\lambda_u/(3.5\lambda_b)]^{-3.5}$ \cite{SMYu}. It provides that density of arbitrary interfering SBSs is equal to $p_a\lambda_b$. Then, active SBSs' coordinates,  consisting a set $\Phi_R(p_a\lambda_b)$, determine dynamics of interference $I^f(t)$.\vskip -7pt
{\small
\begin{equation}
\quad I^f(t)=\sum^{|\Phi_R(p_a\lambda_b)|}_k{ P|h_{k,i}(t)|^2},
\end{equation}} \vskip -10pt
\noindent The signal-to-interference-plus-noise (\textsf{SINR}) with $N_a$ number of transmit antennas is
\vskip -7pt
{\small
\begin{equation}
\mathsf{SINR}(t) = {N_a P |h(t)|^2}/\left({\sigma^2+\frac{\theta_{N_a}}{2\pi}N_aI^f(t)}\right). 
\end{equation}}
\vskip-7pt
\noindent  In the following section, we present the spatially averaged version of $I^f\!(t)$.

\section{Problem Formulation}

The goal of each SBS $k$ is to determine its own caching probability ${p}_{k,j}^*(t)$ for content $j$ in order to minimize a long run average (LRA) cost. The LRA cost is determined by other SBSs' caching policies, states of content request probability, wireless channel, backhaul capacity, and remaining storage size. 
Caching strategies of other SBSs also vary according to the spatio-temporal dynamics. 


\subsection{Interactions: Replication and Interference}

There are inherent interactions among SBSs with respect to their own caching strategies. These interactions depend on replicated contents and interference, which are major bottleneck for optimizing distributed caching. Our purpose is to estimate these interactions in a distributed fashion without full knowledge of other SBSs' states or actions. 

\textbf{Replication.} As shown in Fig. \ref{system_model}, there may be replicated contents downloaded by multiple SBSs within $R$. 
The overlapping contents increase redundant cost due to inefficient resource utilization \cite{social_cost1}. It is worth noticing that the number of replicated contents is determined by other SBSs' caching strategies. 
We define the replication function $I^r_{k,j}(\boldsymbol{p}_{-k,j}(t))$ as the expected value of the overlapping content size per unit storage size $C_{k,j}$.
\vskip -5pt
\begin{equation}
I^r_{k,j}(\boldsymbol{p}_{-k,j}(t))=\frac{1}{C_{k,j}N_{r(j)}}\sum^{|\mathcal{N}|}_{i\neq k} {p}_{i,j}(t), \label{interaction_1}
\end{equation}
\vskip -5pt
\noindent where $\boldsymbol{p}_{-k,j}(t)$ is a vector of caching probabilities of all the other SBSs except SBS $k$, and $N_{r(j)}$ denotes the number of contents whose request probability is asymptotically equal to that of content $j$. Specifically,  $N_{r(j)}$ is the cardinality of a set $\{ m | m \in {M} \text{ such that } |x_m-x_j| \leq \epsilon\}$ where  $\epsilon$ is sufficiently small. 

\textbf{Interference.} Interactions among SBSs through interference can be bottlenecks for optimizing distributed caching.
To incorporate this spatial interaction, we adopt an analysis of \cite{JHP1} on 
interference in  UDN. It provides $\hat{I}^f(t)$, the interference normalized by SBS density and the number of antennas as follows: 
\vskip -10pt
\begin{equation}
\hat{I}^f\!(t)\!=\!(\lambda_u\pi R)^2 N_a^{\!-\frac{1}{2}}\lambda_b^{-\frac{\alpha}{2}}\! \left(\!1\!+\! \frac{1-R^{2-\alpha}}{\alpha-2}\! \right)\!{P}\mathsf{E}_g [|g(t)|^2]. \label{interaction_2}
\end{equation}


\noindent It gives us an average downlink spectral efficiency (SE) $\mathcal{R}_k(t)$ in UDN as follows:

\vskip -15pt
{\small
\begin{equation}
\mathcal{R}_k(t)\!=\!\mathsf{E}_{S,I^f}\!\left[\log(\!1\!+\!\mathsf{SINR}(t))\right] \!\approx \!\mathsf{E}_{S}\log\left(\!\!1\!+\!\frac{S_k(t)}{\frac{\sigma^2}{ N_a \lambda_b^{\alpha /2}} \!+\! \mathsf{E}_{I^f}\![{\hat{I}}^f(t)]}\!\! \right)\!\!, \label{ER_1}
\end{equation}}
\vskip -10pt
\noindent where $\sigma^2$ is the noise power. Eq. \eqref{ER_1}  allows us to investigate the effect of interference on the upper bound of an average SE.

\subsection{Cost Functions}

An instantaneous cost function $J_{k,j}(t)$ defines the LRA cost. It is affected by backhaul capacity, remaining storage size, spectral efficiency, and overlapping contents among SBSs. 
SBS $k$ cannot download more than $B_{k,j}(t)$, the allocated backhaul capacity for downloading content $j$ at time $t$. 
 If $p_{k,j}(t) \!< \!\frac{B_{k,j}(t)}{L_j}$, the backhaul cost $\phi_{k,j}$ is model as $\phi_{k,j}(p_{k,j}(t))\!=\! -\log (B_{k,j}(t)-L_jp_{k,j}(t))$. Otherwise,  the cost function $\phi_{k,j}(p_{k,j}(t))$ goes to infinity, preventing the download rate of content $j$ from exceeding the available backhaul rate. This form of cost function is widely used as in \cite{MFG_caching}.

As cached content files occupy the storage, it causes processing latency \cite{Storage_cost} or delay to search requested files by users.    
We consider this storage cost as follows:
\vskip -10pt
\begin{equation}
\psi_{k,j}(Q_{k,j}(t))= \gamma(C_{k,j}-Q_{k,j}(t))/{C_{k,j}}.  \label{storage cost}
\end{equation}
\vskip -5pt
\noindent Then, the global instantaneous cost is 
{\medmuskip=-1mu\thinmuskip=-1mu\thickmuskip=-1mu
\begin{align}
J_{k,j}(p_{k,j}(t),\boldsymbol{p_{-k,j}}(t))\text{ }&=\text{ }\frac{\phi_{k,j}(p_{k,j}(t))(1\!+\!I^r_{k,j}(\boldsymbol{p}_{-k,j}(t)))}{\mathcal{R}_k(t,I^f(t))x_j(t)}\nonumber\\
&+\text{ }\psi_{k,j}(Q_{k,j}(t)). \label{inst_global_cost}
\end{align}}
\vskip -10pt \noindent With this cost function, we define an average total cost over the long run, called long run average (LRA) cost.

\vskip -15pt

\subsection{Stochastic Differential Game for Edge Caching}

The state of SBS $k$ and content $j$ at time $t$ is defined as 
$\boldsymbol{s}_{k,j}(t)=\{x_j(t),\mathcal{R}_k(t),Q_{k,j}(t)\}$, $\forall k \in \mathcal{N}, \forall j \in \mathcal{M}$. The stochastic differential game (SDG) for edge caching is defined by
$(\mathcal{N},\mathcal{S}_{k,j}, \mathcal{A}_{k,j}, \mathcal{J}_{k,j}    )$ where $\mathcal{S}_{k,j}$ is the state space of SBS $k$ and content $j$, $\mathcal{A}_k$ is the set of all caching controls $\{p_{k,j}(t), 0 \leq t \leq T \}$ admissible for
the state dynamics, and $\mathcal{J}_{k,j}$ is the LRA cost over a time window $[0,T]$ defined as follows:
\vskip -10pt
{\small
\begin{equation}
\mathcal{J}_{k,j} = \mathop{E} \left[\int_t^T J_{k,j}(p_{k,j}(t),\boldsymbol{p_{-k,j}}(t)) \text{ d}t + \kappa(\underline{\boldsymbol{s}}_{k}(T))\right],
\end{equation}}
\vskip -10pt
\noindent where $\kappa: \mathcal{S}_{k,j} \rightarrow \mathbb{R}, \underline{\boldsymbol{s}}_{k} \mapsto \kappa(\underline{\boldsymbol{s}}_{k})$ is the cost of having a remaining storage size at the end of the time window $[0,T]$. Thus, the SDG is formulated as follows:


\vskip -7pt
\begin{equation}
\hspace{-70pt} \textbf{(P1)} \qquad\qquad\quad\quad v_{k,j}(t)=\mathop{\text{inf}}\limits_{p_{k,j}(t)}\text{ } \mathcal{J}_{k,j}(t).
\end{equation}
\vskip -16pt
\begin{flalign}
\text{subject to }\quad\quad&\text{d}x_j(t)= (u_j-a_j)\text{d}t+\eta \text{d}W_j(t), \quad\label{const_1} \\ 
&\text{d}Q_{k,j}(t)=(\mu_{k,j}-L_j p_{k,j}(t))\text{d}t.   \label{const_2}
\end{flalign} 

The existence of a Nash equilibrium of problem \textbf{P1} is guaranteed if there exists a joint solution of the
following coupled HJB equations for all $k$ and $j$ \cite{exist_HJBsol1}: 
\vskip -10pt
{\small
\begin{align}
0&\!= \partial_t v_{k,j}(t)\! +\!\mathop{\text{inf}}\limits_{p_{k,j}(t)} \bigg{[}J_{k,j}(p_{k,j}(t),\boldsymbol{p_{-k,j}}(t))\!+\! \frac{\eta^2}{2}\partial_{xx}^2 v_{k,j}(t)\nonumber \\
 &+\! \underbrace{(\mu_k -L_j p_{k,j}(t))}_{(A)} \partial_{Q_k}v_{k,j}(t)+\underbrace{(u_j\!-\!a_j)}_{(B)}\partial_x v_{k,j}(t)\bigg{]}. \label{hjb_sdg}
\end{align}}
\vskip -10pt
If the smoothness of the drift functions (A) and (B) in the dynamic equation \eqref{hjb_sdg} and the cost function \eqref{inst_global_cost}, we can assure that 
a unique solution of the equation \eqref{hjb_sdg} exists \cite{exist_HJBsol1}.
Unfortunately, it is complex to solve the coupled $N \times M$ HJB equations.
We thus release complexity of this system in the next section.


\begin{figure}
\centering
\includegraphics[angle=0, height=0.4\textwidth]{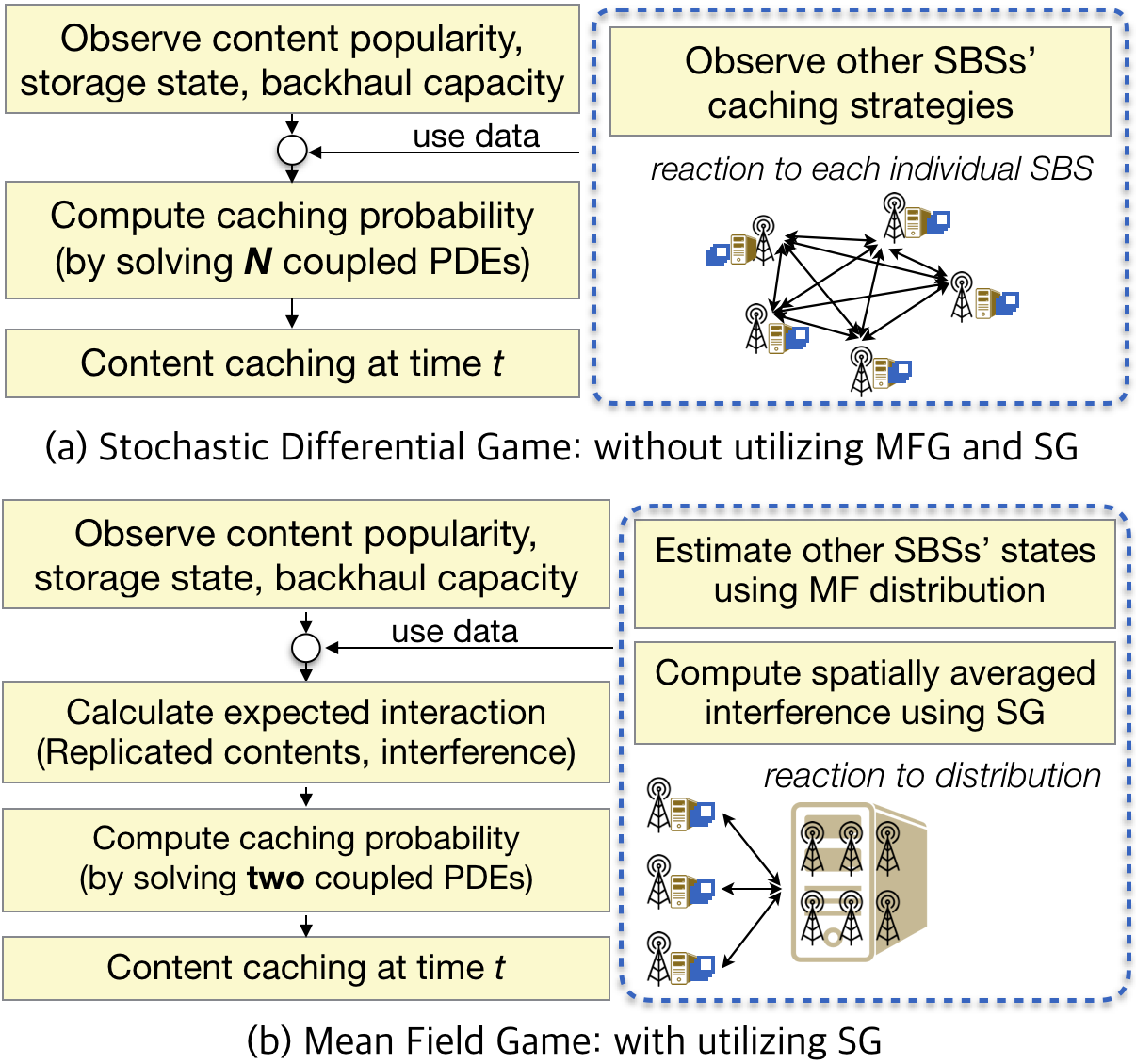}   
\caption{\small{Ultra-dense edge caching flow charts with and without utilizing MFG and SG.}  }\label{diagram} \vskip-10pt
\end{figure}


\subsection{Mean-field Game for Caching}

It is hard to react to each individual SBS's caching strategy in solving the SDG \textbf{P1}.
Mean-field game (MFG) theory enables us to transform these multiple interactions into a single interaction, called MF interaction, via MF approximation. 
According to \cite{MF_ref2}, this approximation holds under the following conditions: 
a large number of players, excahngeability of players under the caching control and finite MF interaction.

Remark that the first condition corresponds to the definition of UDNs.
Players in the SDG are said to be exchangeable or indistinguishable under the control ${p}_{k,j}(t)$ and the states of players and contents, 
if the player's control is invariant by their indices and decided by only their individual states.
In other words, permuting players' indices cannot change their control policies. Under this exchageability, we can focus on a generic SBS by dropping its index~$k$. 

Respective MF interactions \eqref{interaction_1} and \eqref{interaction_2} should asymptotically converge to a finite value under the above conditions.
MF replication \eqref{interaction_1} goes to zero when the number of contents per SBS is extremely large, i.e. $M \gg N $. Such a condition implies that the cardinality of the set consisting of asymptotically equal content popularity goes to infinity. In other words, $N_{r(j)}$ goes to infinity yielding that $I^r_{k,j}(\boldsymbol{p}_{-k,j}(t))$ becomes zero. 
In terms of interference, MF interference converges  as
the ratio of SBS density to user density goes to infinity, i.e.  $N_a\lambda_b^{\alpha}/(\lambda_u R)^4 \rightarrow \infty$ \cite{JHP1}. Such a condition corresponds to the notion of UDN  \cite{interf_udn} or massive MIMO ($N_a \rightarrow \infty$). That is, the conditions enabling MF approximation  to inherently hold under ultra-densified caching networks.

To approximate interactions from other SBSs, we need a state distribution of SBSs and contents at time $t$, called MF distribution $m_t(x,Q)$.
It is defined by a counting measure $M_t^{(N\times M)}(x,Q)= \frac{1}{NM}\sum_{j=1}^{M}\sum_{k=1}^{N} \delta_{\{x_j(t),Q_k(t)\}}$.
\noindent We assume that the empirical distribution $M_t^{(N \times M)}(x,Q)$ converges to $m_t(x,Q)$, which is the density of contents and SBSs in state $(x,Q)$. Note that we omit the SE $\mathcal{R}(t)$ from the density measure to only consider temporally correlated state without loss of generality.



The MF distribution $m_t(x,Q)$  is a solution of the following Fokker-Planck-Kolmogorov (FPK) equation. 
\begin{align}
0&= \partial_t m_t(x,Q) -(u_j-a_j)\partial_x m_t(x,Q) \nonumber \\&+ (\mu_k -L_j p_{j}(t)) \partial_{Q}m_t(x,Q) 
 + \frac{\eta^2}{2}\partial_{xx}^2 m_t(x,Q).  \label{fpk_1}
\end{align}

\noindent Let us denote the solution of the FPK equation \eqref{fpk_1} as $m_t^*(x_j,Q)$. 
Exchangeability and existence of the MF distribution allow us to approximate the interaction $I^r_{k,j}(\boldsymbol{p}_{-k,j}(t))$ as a function of $m_t^*(x_j,Q)$ as follows:
\vskip-8pt
{\small\begin{equation}
I^r_{j}(t,m_t^*(x_j,Q))=\int_{\!Q}\!\int_{\!x} \frac{m_t^*(x_j,Q)}{C_{k,j}N_{r(j)}} {p}_{j}(t,x,Q)\text{d}x\text{d}Q, \label{MF_interaction_apprx}
\end{equation}}
\vskip-8pt
\noindent Remark that we can estimate the interaction from replication \eqref{MF_interaction_apprx} without observing other SBSs' caching strategies.
Thus, it is not necessary to have full knowledge of the states or the caching control policies of other SBSs. It provides that an SBS only need to solve a pair of equations, the FPK equation \eqref{fpk_1} and the following modified HJB one from \eqref{hjb_sdg}:  
\vskip -10pt
{\small  
\begin{align}
0&\!= \partial_t v_{j}(t)\! +\!\mathop{\text{inf}}\limits_{p_{j}(t)} \bigg{[}J_{j}(p_{j}(t),I_j(t,m_t^*(x,Q))\!+\! \frac{\eta^2}{2}\partial_{xx}^2 v_{j}(t)\nonumber \\
 &+\! (\mu -L_j p_{j}(t)) \partial_{Q}v_{j}(t)+(u_j\!-\!a_j)\partial_x v_{j}(t)\bigg{]}. \label{hjb_mfg}
\end{align}}
\vskip -10pt
\noindent We can solve this equation in backward and find the solution of the stochastic optimization problem \textbf{P1}. It is worth mentioning that the number of PDEs to solve for one content is reduced to two from $N$. The respective processes of solving \textbf{P1} in ways of SDG and MFG are depicted in Fig. \ref{diagram}.

\vskip 10pt
 \noindent {\bf Proposition 1.} {\it The optimal caching probability
is given by: }
\begin{eqnarray}
p_{j}^*(t)=\frac{1}{L_j}\left[B_{j}(t)- \frac{1+I^r_j(t,m^*_t(x_j,Q))}{\mathcal{R}(t,I^f(t)) x_j(t)\partial{\scriptscriptstyle 
 {Q}}{v^*_{j}}}    \right]^+,  \label{Propo1}
 \end{eqnarray}
 {\it where $m_t^*(x,Q)$ and $v_j^*(t)$ are the unique solutions of  \eqref{fpk_1} and \eqref{hjb_mfg}, respectively. }  
 
\noindent {\it Proof}: Appendix. 
\vskip 5pt
\begin{figure}
\centering
\includegraphics[angle=0, height=0.25\textwidth]{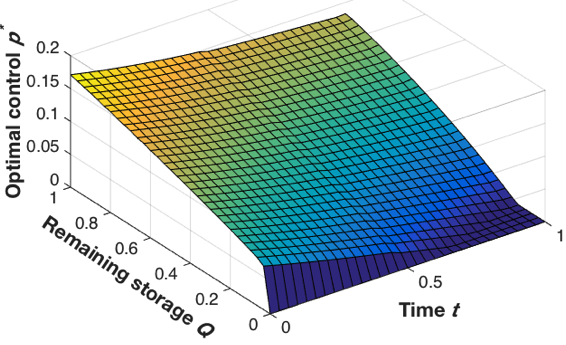}   
\caption{\small{Evolution of the optimal caching control $p^*(t,Q)$ at the MF equilibrium when $x_j=0.4$. The initial MF distribution $m_0$ is given as $\mathcal{N}(0.7,0.05^2)$. }}\label{trajectory_control} \vskip -10pt
\end{figure}

The optimal caching probability $p_{j}^*(t)$ is calculated by water-filling fashion, where the allocated backhaul capacity $B_j(t)$ determines the water level. Noting that SE $\mathcal{R}(t)$ increases with the number of antenna $N_a$ and SBS density $\lambda_b$, SBSs cache more contents from server when wireless capacity is improved. The partial derivative $\partial{\scriptscriptstyle{Q}}{v_{j}}$ implies the effect of the current decision on the final LRA cost.

\begin{figure*}
\centering
\includegraphics[angle=0, width=0.94\textwidth]{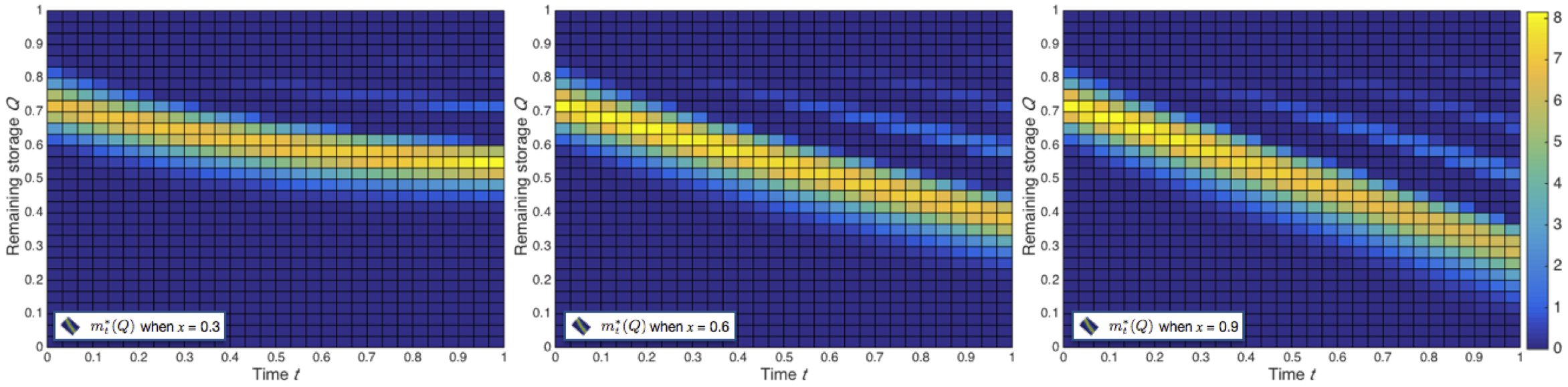} 
\caption{{\small Evolution of the MF distribution $m^*_t(Q)$ that indicates the density of SBSs having the remaining storage space $Q$ at time $t$ with respect to different content popularity $x,$ $B=1, N_{r(j)}=20, \mu =0.1, \lambda_u=0.001,\lambda_b=0.03,R=10/\!\sqrt{\pi}$.}}\label{MF_dist_total} 
\end{figure*}

The respective solutions $m_t^*(x,Q)$ and $v_j^*(t)$ of \eqref{fpk_1} and \eqref{hjb_mfg} define the {\it mean field equilibrium} (MFE) which corresponds to the notion of  Nash equilibrium in the MFG framework \cite{MFG_application}. 
Note that existence and uniqueness of the optimal caching control is guaranteed. It verifies that the optimal caching algorithm converges to a unique MFE, when the initial conditions $m_0$, $x_j(0)$, and $Q(0)$ are given. 

The optimal caching algorithm in Proposition 1 becomes more tractable when each SBS affords high storage capacity enough  to ignore cost for storage usage.
Thus, we can obtain the following corollary of Proposition 1.


\vskip 5pt
 \noindent {\bf Corollary 1.} {\it When the storage capacity is sufficiently high ($C_{k,j} \rightarrow \infty$), the optimal caching control is given by}
\begin{eqnarray}
p^*_{j}(t)=\frac{1}{L_j}\left[B_{j}(t)- \frac{1+I^r_j(t,m^*_t(x_j,Q))}{\mathcal{R}(t,I^f(t)) x_j(t)\hat{\gamma}}    \right]^+,
\end{eqnarray}                                            \label{Coroll1}
\vskip -10pt
\noindent 
where $\hat{\gamma}$ is a constant.
\vskip 2pt
 \noindent{\it Proof}: When the storage capacity goes to infinity, the instantaneous cost function for storage usage \eqref{storage cost} is given as follows: 
$ \lim_{C_{j}\rightarrow \infty} \psi_{j}(Q_{j}(t))= \gamma.$
Hence, the partial derivative of LRA cost with respect to the remaining storage size $\partial{\scriptscriptstyle{Q}}{v_{j}}$ also has a constant value of $\hat{\gamma}$. \hfill$\blacksquare$

It is worth noticing that the optimal caching control becomes independent from the LRA cost dynamics with respect to the storage state.
\vskip -10pt


\section{Numerical Results}

In this section, we present numerical results  for cases in which content request probability remains static and temporally varies according to its evolution law \eqref{SDE_x}.
Let us assume full knowledge of contents request probability and their evolution laws. The initial distribution of the SBSs $m_0$ is given as a normal distribution $\mathcal{N}(0.7,0.05^2)$. 
We assume that the storage size $Q$ belongs to a set $[0,1]$. Assuming Rayleigh fading with mean one, we set the parameters as follows: $\mu=0.1, \gamma=0.01,\lambda_u=0.001,\lambda_b=0.03,R=10/\!\sqrt{\pi}, B=1,N_{r(j)}=20, Q(0)=0.7, \eta=0.1,\alpha=4$.
 
\vskip -10pt
\subsection {Static content request probablity}

We numerically analyze the algorithm performance when the content request probability is static. 
In this case, the problem becomes simplified, because the caching control strategies does not depend on the evolution law of the content popularity. 
Specifically, in HJB \eqref{hjb_mfg}  and FPK \eqref{fpk_1} equations, the derivative terms with respect to content request probability $x$ become zero.


\begin{figure}
\centering
\includegraphics[angle=0, height=0.28\textwidth]{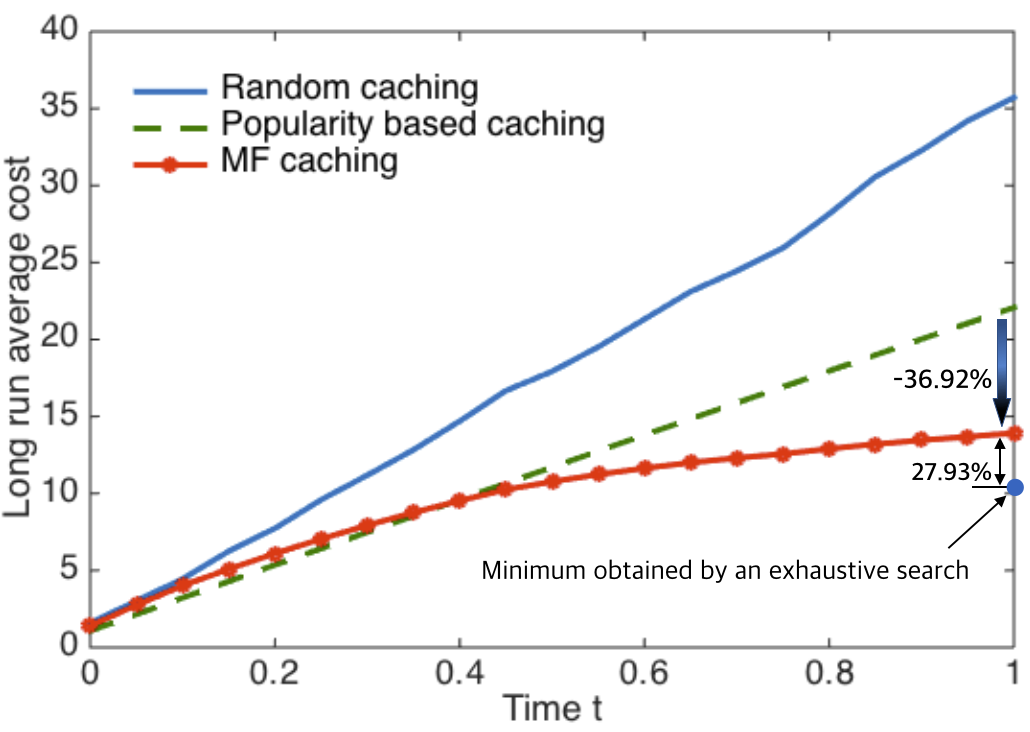}   
\caption{{\small Long run average costs of different caching strategies ($Q(0)=0.7, x(0) =0.3, \eta=0.1, a=0.15,u=0.1$).} }\label{LRA} \vskip -15pt
\end{figure}

Fig. \ref{trajectory_control} shows the evolution of the optimal caching probability $p^*(t)$ for a content with respect to storage state. 
We observe that the value of $p^*(t)$ is lower than the content request probability at all time slots and decreases when available storage space $Q$ becomes small. The reason  is that SBSs restrain replication of the content and save storage capacity for downloading more popular contents.

Fig. \ref{MF_dist_total}  represents evolutions of MF distribution $m_t^*(Q)$ for a content with different popularity. Note that $m_t^*(Q)$ indicates the density of SBSs whose unoccupied storage size is equal to $Q$ at time $t$. 
We observe that the unoccupied storage space of SBSs does not diverge from each other. It validates that the proposed algorithm achieves the MFE.
In this equilibrium, the amount of downloaded content file decreases when the content popularity $x$ becomes low. This tendency corresponds to the trajectory of the optimal caching probability in Fig. \ref{trajectory_control}.
Almost every SBS has downloaded the content over time, but not spent its whole storage. The remaining storage saturates even though the content popularity is equal to $0.9$. 
It implies that SBSs reduce the  downloading amount of the popular content in consideration of the expected replication amount, which increases with the popularity.


%


\begin{figure}
\centering
\includegraphics[angle=0, height=0.285\textwidth]{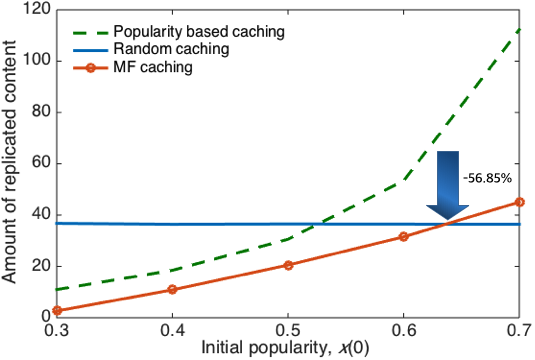}   
\caption{{\small The amount of replicated contents per storage usage ($Q(0)=0.7, \eta=0.1, a=0.15,u=0.1$). } }\label{Repli} \vskip -15pt
\end{figure}


\subsection {Dynamic content request probablity}

We investigate the performance of the proposed MF caching algorithm under temporal dynamics of the content popularity \eqref{SDE_x}. 
Note that the temporal dynamics constrains the optimal caching strategy to follow the optimal trajectory $[v_j^*(t),m_t^*(x,Q)]$, a set of solutions to the coupled equations \eqref{hjb_mfg} and \eqref{fpk_1}. 

Fig.~\ref{LRA} shows  the LRA cost of the proposed MF caching algorithm
compared to results of a popularity based algorithm and random caching one. The blue dot in the figure indicates the minimal LRA cost obtained by an exhaustive search. 
A slight disagreement between the proposed and exhaustive algorithm by full search appears from two reasons; one is that each segment size of time $t$ and storage capacity $Q$ is not infinitesimal, leading to inaccurate partial derivative values, and the other is the finite number of SBSs. Nevertheless, the proposed caching strategy reduces about $36\%$ of the LRA cost as compared to an algorithm that determines caching probability in proportional to the request probability. 
This performance gain results from allowing the system to avoid the expected replication of contents and interference varying according to the spatio-temporal dynamics.

Fig. \ref{Repli} shows the amount of replicated contents per storage usage as a function of initial content probability $x(0)$. The proposed MF caching algorithm reduces caching content replication averagely 56\% compared to popularity based caching. However, MF caching algorithm yields the higher amount of replication than random caching does when the content request probability becomes high. The reason is that the random policy downloads contents regardless of their own popularity, so the amount of replication remains steady. On the other hand, MF caching increases the downloaded volume of popular contents.

\section{Concluding Remarks}

In this paper, we proposed an edge caching algorithm for ultra-dense edge caching networks (UDCNs), taking into account the spatio-temporal user demand and network dynamics. 
Leveraging the frameworks of MFG theory and SG, our algorithm enables SBSs to distributively determine their own caching strategy without full knowledge of other SBSs' caching strategies or storage state. The proposed MF caching algorithm reduces not only the long run average cost but also the replicated content amount compared to a caching control that is merely based on content popularity.
Future work can extend to an optimal caching strategy to cope with imperfect  knowledge of content popularity dynamics.

\section{Appendix: Proof of Proposition 1}
The optimal control strategy is the argument of the infimum term of the HJB equation \eqref{hjb_sdg}.
\vskip -10pt
{\small
\begin{align}
\mathop{\text{inf}}\limits_{p_{j}(t)} \bigg{[}&J_{j}(p_{j}(t),I^r_j(t,m_t^*(x,Q))\!+\! \frac{\eta^2}{2}\partial_{xx}^2 v_{j}(t)\nonumber \\
 &+\! (\mu -L_j p_{j}(t)) \partial_{Q}v_{j}(t)+(u_j\!-\!a_j)\partial_x v_{j}(t)\bigg{]} \label{infimum}
\end{align}} 
\vskip -10pt

\noindent Solving \eqref{infimum} for all time $t$ is immediate by convexity of the objective function, of which the critical point is the unique optimal control $p_j^*(t)$ calculated as described in \eqref{Propo1}.
Note that $p^*_j(t)$ is a function of the solutions of the equations \eqref{fpk_1} and \eqref{hjb_mfg}.
The expression of $p_j^*(t)$ \eqref{Propo1} provides the final versions of the HJB and FPK equations composing MFG as follows:
\vskip -10pt
{\small
\begin{align}
0&= \partial_t v_{j}(t) -\frac{\log \left(B_{j}(t)-\left[B_{j}(t)- \frac{1+I^r_j(t,m_t^*(x,Q))}{\mathcal{R}(t,I^f(t))x_j(t)\partial{\scriptscriptstyle{Q}}{v_{j}}}    \right]^{\scriptscriptstyle{+}}\right)}{\mathcal{R}(t,I^f(t))x_j(t)} \nonumber \\
 &\!\times \!(1\!+\!I^r_j(t,m_t^*(x,Q)))+\!\frac{\alpha(C_j\!-\!Q_j(t))}{C_j}+(u_j\!-\!a_j)\partial_x v_{j}(t)\nonumber \\
 &\!+\! \left(\!\mu \!-\!\left[\!B_{j}(t)\!- \!\frac{\!1\!+\!I^r_j(t,m_t^*(x,\!Q))}{\mathcal{R}(t,I^f(t))x_j(t)\partial{\scriptscriptstyle{Q}}{v_{j}}}\!\right]^{\!\scriptscriptstyle{+}}\right) \!\partial_{Q}v_{j}(t)\!+\! \frac{\eta^2}{2}\partial_{xx}^2 v_{j}(t), \nonumber\\
0&= \partial_t m_t(x,Q) -(u_j-a_j)\partial_x m_t(x,Q) + \frac{\eta^2}{2}\partial_{xx}^2 m_t(x,Q)
\nonumber \\&+ \left(\mu_k -\left[\!B_{j}(t)\!- \!\frac{\!1\!+\!I^r_j(t,m_t(x,\!Q))}{\mathcal{R}(t,I^f(t))x_j(t)\partial{\scriptscriptstyle{Q}}{v_{j}^*}}\!\right]^{\!\scriptscriptstyle{+}}\right) \partial_{Q}m_t(x,Q). \nonumber
\end{align}}
\vskip -10pt
\noindent From these equations, we can find the values of
$v_j^*(t)$ and $m_t^*(x,Q)$. Note that the smoothness of the drift functions and in the dynamic equation and the cost function \eqref{inst_global_cost} assures the uniqueness of the solution \cite{exist_HJBsol1}. \hfill$\blacksquare$

\section*{Acknowledgement}

This research was supported by the Korean government through the National Research Foundation (NRF) of Korea Grant (NRF-2014R1A2A1A11053234), the Institute for Information \& communications Technology Promotion (IITP) grant funded by the Korea government (MSIP) (No. 2015-0-00294, Spectrum Sensing and Future Radio Communication Platforms), the Academy of Finland CARMA project, the FOGGY project and the ERC Starting Grant 305123 MORE
(Advanced Mathematical Tools for Complex Network Engineering).


%

\ifCLASSOPTIONcaptionsoff
  \newpage
\fi

\end{document}